# Ferroelectricity driven by magnetism in quasi-one-dimensional Ba$_9$Fe$_3$Se$_{15}$

Jun Zhang[†1, 2], Xiancheng Wang[†,*1,2], Yiqing Hao[†3], Guangxiu Liu[†1,2], Long Zhou[1], Daniel M. Pajerowski[4], Jian-Tao Wang[1,2,5], Jinlong Zhu[6], Jun Zhao[3,7], Jiulong Wang[8], Yifeng Zhao[8], Chungang Duan*[8], Youwen Long*[1,2,5], Chang-Jong Kang[9], Martha Greenblatt[10], and Changqing Jin*[1,2,5]

**Abstract:** The spin-induced ferroelectricity in quasi-1D spin chain system is little known, which could be fundamentally different from those in three-dimensional (3D) system. Here, we report the ferroelectricity driven by a tilted screw spin order and its exotic dynamic in the spin-chain compound Ba$_9$Fe$_3$Se$_{15}$. It is found that the spin-induced polarization has already occurred and exhibits magnetoelectric coupling behavior far above the long-range spin order (LRSO) at $T_N$ = 14 K. The polarized entities grow and their dynamic responses slow down gradually with decreasing temperature and permeate the whole lattice to form 3D ferroelectricity at $T_N$. Our results reveal that the short-range spin orders (SRSOs) in the decoupled chains play a key role for the exotic dynamic in this dimension reduced system. Ba$_9$Fe$_3$Se$_{15}$ is the only example so far which exhibits electric polarization above LRSO temperature because of the formation of SRSOs.

[1]Beijing National Laboratory for Condensed Matter Physics, Institute of Physics, Chinese Academy of Sciences, Beijing 100190, China. [2]School of Physics, University of Chinese Academy of Sciences, Beijing 100190, China. [3]State Key Laboratory of Surface Physics and Department of Physics, Fudan University, Shanghai 200433, China. [4]Neutron Scattering Division, Oak Ridge National Laboratory, Oak Ridge, Tennessee 37831, USA. [5]Materials Research Lab at Songshan Lake, Dongguan 523808, China. [6]Department of Physics, Southern University of Science and Technology, Shenzhen 518055, China. [7]Institute of Nanoelectronics and Quantum Computing, Fudan University, Shanghai 200433, China. [8]Key Laboratory of Polar Materials and Devices, East China Normal University, Shanghai 200062, China. [9]Department of Physics, Chungnam National University, Daejeon 34134, South Korea. [10]Department of Chemistry and Chemical Biology, Rutgers, the State University of New Jersey, Piscataway, New Jersey 08854, United States. †These authors contributed equally. *Corresponding author: wangxiancheng@iphy.ac.cn; cgduan@clpm.ecnu.edu.cn; ywlong@iphy.ac.cn; jin@iphy.ac.cn

## Introduction

The multiferroic compounds, where the ferroelectricity (FE) is generated by special spin configuration, have received much attention due to their interesting physics and potential applications[1-5]. Many special spin ordering arrangements can break the space-reversal symmetry and allow the occurrence of spontaneous FE, such as the typical example of spin spiral orders[1,2]. For the mechanism of FE induced by magnetism in spiral spin arrangements, the spin current model is proposed, where spiral spins can induce instantaneous spin current and lead to electric polarization ($P$) accordingly[6]. An alternative explanation of $P$ driven by spiral order is the inverse Dzyaloshinskii-Moriya (DM) interaction[7]. That is, the spiral spins can displace the ligand ions via the electron-lattice interaction to generate $P$. In these models, the polarization $P$ generated by two spin moments can be described with $\vec{P} \propto \vec{e}_{ij} \times (\vec{S}_i \times \vec{S}_j)$, where $\vec{e}_{ij}$ is the unit vector connecting the sites of $\vec{S}_i$ and $\vec{S}_j$. Various spiral magnets have been studied and their polarization behavior follows the spin-current model[8-16], such as $TbMnO_3$[8] and $ZnCr_2Se_4$[11]. $TbMnO_3$ is a typical magnet with cycloidal spiral structure and exhibits FE, where the spin rotation axis ($\vec{S}_i \times \vec{S}_j$) is perpendicular to the spin chain[8,17]. Besides the cycloidal structure, screw structure is another spin spiral configuration, where the spin rotation axis is parallel to the spin chain. In this case $P$ should be zero according to the spin-current model. $ZnCr_2Se_4$ was reported to adopt such a proper screw-type spin order. Applying magnetic field $H$ can transform the screw structure into a conical spin structure with the cone axis being oriented with $H$ [18]. When the cone axis is tilted from the direction of $\vec{e}_{ij}$, the conical spins can generate non-zero $P$, as studied in $ZnCr_2Se_4$ [11].

For most compounds with spin-driven polarization, such as the three-dimensional (3D) $R$MnO$_3$[8], $MnWO_4$[12], $La/BiMn_3Cr_4O_{12}$[5,19] and two-dimensional (2D) $CuFeO_2$[9], the dielectric permittivity is frequency independent and displays a sharp peak corresponding to the magnetic ordering transition. In fact, the sharp peak is a common phenomenon for ferroelectrics, because the dipoles will be spontaneously polarized in a very narrow temperature window when the ferroelectric transition occurs. However, the dynamic properties of quasi one-dimensional (1D) spin systems are anticipated to be different, due to the existence of abundant short-range spin orders (SRSOs) in the decoupled spin chains above the long-range spin ordering (LRSO) transition temperature[20, 21]. It is expected for the SRSOs to generate nano-size polarized domains and respond to the applied AC electric field and thus, lead to exotic physical phenomena. Although $LiCu_2O_2$ and $Ca_3(Co,Mn)_2O_6$ were reported to have quasi-1D spin chains and show spin-driven ferroelectricity, their dielectric permittivity exhibit the similar sharp dielectric peak with 2D/3D multiferroics[4, 14].

Here, we report the studies about the spin induced polarization and its dynamic in the newly discovered compound $Ba_9Fe_3Se_{15}$ with strong quasi-1D Fe-chains[22], which crystallizes in a hexagonal $Hf_5Sn_3Cu$ anti-type structure[23-29]. $Ba_9Fe_3Se_{15}$ undergoes a spiral magnetic ordering at $T_N$ ~14 K which is accompanied by the occurrence of spontaneous FE. Interestingly, the FE occurs slightly higher than $T_N$ and the dielectric relaxation is observed above $T_N$, which are caused by the formation of SRSOs due to the quasi 1D spin chain characteristic. Thus, $Ba_9Fe_3Se_{15}$ is a unique system with 1D chains and provides a good example to study the microscopic mechanisms of the spin-induced electric polarization and dielectric relaxation with magnetic origin.

## Results

The structure of newly discovered $Ba_9Fe_3Se_{15}$ was solved combining powder and single crystal X-ray diffraction (XRD)[22]. It crystallizes in a monoclinic crystal structure with the space group *C2/c* and lattice parameters $a$ = 16.5947 Å, $b$ = 9.6128 Å, $c$ = 18.6735 Å and $\beta$ = 90.11 °. The crystal structure consists of face-sharing $FeSe_6$ octahedral chains running along the *c* axis and separated with a large distance of ~9.6 Å by $Ba^{2+}$ ions and Se-chains, thus exhibiting strong 1D spin-chain structure (Fig. 1a). There are two Fe sites in the $FeSe_6$ chain, and the oxidation state of Fe is 2+ for each site[22].

Fig. 1b shows the DC magnetization for single crystalline $Ba_9Fe_3Se_{15}$ measured by the applied field $H//ab$-plane. The magnetization curves increase sharply at ~14 K, showing a ferromagnet-like transition. The shape of the magnetization with $H//c$-axis is similar to and one order of magnitude smaller than that with $H//ab$-plane, as shown in Fig. S1(a). The isothermal $M(H)$ for $H//ab$-plane and $H//c$-axis at 2 K is shown in Fig. S1(b). For $H//ab$, the magnetization exhibits ferromagnet-like behavior with a weak saturation field of ~400 Oe. In contrast, the magnetization basically exhibits a linear field dependence for $H//c$. In the case of $H//c$ in Fig. S1(a,b), the small ferromagnetic component is most likely the result of a misalignment of the sample by a few degrees from the $c$-axis. The saturation magnetic moment at 2 K and 7 T is about 1.06 $\mu_B$/Fe, which is much smaller than the expected magnetic moment of $Fe^{2+}$ with S = 2. This observation indicates that few of the spins of Fe are oriented along the field. With increasing $H$, the magnetization continues to increase to ~2.7 $\mu_B$/Fe at $H$ = 60 T (Fig. S1(c)). These results indicate that the magnetic moments in $Ba_9Fe_3Se_{15}$ are aligned within the $ab$-plane.

Fig. 1(c) shows the specific heat data of $Ba_9Fe_3Se_{15}$. Although the magnetization shows a sharp increase corresponding to a LRSO transition, only a small kink is observed in the specific heat at the LRSO temperature, which can only be clearly seen as an anomaly in the derivative of specific heat at ~14 K. This is a typical phenomenon for the system with quasi-1D spin chains. In an ideal 1D spin chain, the intrachain coupling cannot lead to the formation of LRSO at finite temperature, because of the strong quantum or thermal fluctuations in the reduced dimension[20]. However, for quasi-1D spin chains, the interchain exchange interaction plays a key role in the LRSO formation although it is much weaker than the intrachain coupling. As a result, short-range spin correlations related with intrachain coupling have already developed before the LRSO transition, and leads to the release of most of the magnetic entropy. Therefore, the small kink observed in the specific heat data corresponding to the LRSO transition corroborates that $Ba_9Fe_3Se_{15}$ has a strong quasi-1D character.

Neutron powder diffraction (NPD) measurements have been performed to investigate the magnetic structure of $Ba_9Fe_3Se_{15}$. Fig. S2 shows the NPD pattern at 4

K and 40 K. The main difference between these two patterns is the appearance of new magnetic Bragg peaks at 4 K. The temperature dependence of integrated intensity for the magnetic Bragg peak (1 1 2)-Q is shown in Fig. 1(d). A clear magnetic phase transition is observed at $T_N$ ~14 K, which agrees with that observed by the susceptibility and specific heat experiments. Fig. 2(a) shows the Rietveld fitting of the NPD data. The magnetic diffraction peaks are observed in incommensurate reciprocal lattice positions, which were well indexed as (H,K,L)±q with the magnetic wave vector q = (0 0 0.130(4)), giving the *c*-axis magnetic lattice parameter ~143.6 Å. The correlation length along the *c*-direction is of the order of $\xi_c$ ~ 35(1) Å, which is just nearly two times that of the *c* axis dimension (~18.6 Å). In addition, compared with the fully ordered moment expected for $Fe^{2+}$ spin, the obtained ordered magnetic moment is reduced to only 2.72(6) $\mu_B$/Fe. This should be attributed to strong quantum or thermal fluctuations due to the reduced dimensions. Fig. 2(b) shows the magnetic structure of $Ba_9Fe_3Se_{15}$, which displays a helical magnetic structure whose magnetic moments is spiral in the *ab*-plane. This in-plane spin alignment agrees with the result of the magnetic susceptibility measurements. Since there are two Fe sites, one would not expect a realization of a perfect helical magnetic structure. The angle between neighboring spins of Fe1 and Fe2 atoms is ~116.1° while the angle between neighboring spins of Fe2 atoms is ~104.3°. Because $Ba_9Fe_3Se_{15}$ is monoclinic, the spin chain along the *c*-axis is not perpendicular to the spin rotation plane (*ab*-plane), which means that the spiral ordering is a slightly tilted screw-type one.

Since a spiral magnetic structure usually generates electric polarization, we performed the polarization measurements for a polycrystalline sample of $Ba_9Fe_3Se_{15}$. After poling the sample from 20 K to 2 K under a zero magnetic field, the pyroelectric current $I_p$ was measured during the warming process, and the polarization $P$ can be obtained by integrating $I_p$, as shown in Fig. 3(a, b). After the pyroelectric current $I_p$ reaches to the maximum value, it gradually decays to zero at ~16 K (marked by the blue dashed line) and the integration of $I_p(T)$ gives an electric polarization $P$ about 6 $\mu C\text{-}m^{-2}$. In addition, the sign of $I_p$ and $P$ can be reversed symmetrically by changing the sign of poling electric field. To compare the transition temperatures between polarization and LRSO, the temperature derivative of specific heat is re-plotted (Fig.3(c)). The violet dashed line represents the LRSO transition temperature (14 K). The occurrence of polarization is very close to the LRSO formation, slightly higher by 2 K, which suggests that the polarization is associated with the magnetism in $Ba_9Fe_3Se_{15}$. Comparing with the spiral magnets, such as $TbMnO_3$ ($P$~800 $\mu C\text{-}m^{-2}$) [8] and $ZnCr_2Se_4$ (maximum $P$~60 $\mu C\text{-}m^{-2}$) [11], the polarization value of $Ba_9Fe_3Se_{15}$ is very small (6 $\mu C\text{-}m^{-2}$); it should arise mainly from the small angle between the $c$ axis and the spin rotation axis according to the prediction from spin current model with $\vec{P} \propto \vec{e}_{ij} \times (\vec{S}_i \times \vec{S}_j)$. Moreover, our $P$ measurement under different $H$ indicates that the value of $P$ can be increased by ~25% by applying $H$ = 9 T, as shown in Fig. 3(d). The variation of the polarization under magnetic field should be caused by the formation of conical spin structure, which has been studied in $ZnCr_2Se_4$ [11].

The temperature dependence of dielectric permittivity $\varepsilon_r$ as well as dielectric loss (tan$\delta$) measured under different frequencies and zero magnetic field are shown in Fig. 4(a, b). The $\varepsilon_r(T)$ curve shows a broad hump and drops quickly around the LRSO temperature $T_N$, and the tan$\delta$ exhibits a clear peak. Both $\varepsilon_r$ and tan$\delta$ are frequency dependent, and the peaks of tan$\delta$ shift toward high temperature with the increase of frequency, demonstrating a typical dielectric relaxation behavior. The small value of tan$\delta$ (< 0.05) indicates that the dielectric loss comes from the intrinsic response to the applied AC electric field $E$. The dielectric relaxation is usually observed in $ABO_3$ perovskites with disordered lattice[30-34]. In these systems the dipole hopping between equivalent dipolar orientations is thermally activated and thus, the relaxation time $\tau$ (or 1/$f$) obeys the Arrhenius law, $f = f_0 \exp(-\frac{\Delta}{k_b T})$. For $Ba_9Fe_3Se_{15}$, Fig. 4(c) shows the relationship between the frequency and the peak temperature of tan$\delta$. The linear correlation between Ln($f$) and inverse temperature suggests that the relaxation in $Ba_9Fe_3Se_{15}$ is also a thermally activated process. After fitting the data, we can obtain the activation energy about 31 meV.

The $\varepsilon_r$ and tan$\delta$ data measured under different magnetic field show similar dielectric relaxation behaviors with varying temperature and frequency (Fig. S3(a-h)).

Fig. 5(a, b) displays the $\varepsilon_r$ and tan$\delta$ data measured at fixed frequency of $10^6$ Hz and different magnetic field, and the inset shows the enlarged view. It is clearly seen that $\varepsilon_r$ is reduced and the tan$\delta$ curves are shifted to high temperature when the magnetic field increases, which is indicative of magnetoelectric coupling far above $T_N$ since SRSOs can be tuned by magnetic field. This varying of $\varepsilon_r$ and tan$\delta$ with magnetic field further confirms the intrinsic dielectric relaxation behavior, and rules out the possibility that the dielectric loss arises from trapped charge carriers. Fig. 5(c) shows the relationship between Ln(*f*) and the peak temperature of tan$\delta$ for different magnetic field, from which different activation energy gaps can be obtained. The energy gap increases almost linearly with applied magnetic field, as shown in the inset of Fig. 5(c).

## Discussions

Ba$_9$Fe$_3$Se$_{15}$ crystallizes in a monoclinic structure and consists of strong quasi-1D spin chains. In the magnetic ordering state, the spins rotate within the *ab*-plane, forming a tilted screw-type spin ordering. According to the spin-current theory, the spin-induced polarization *P* can be given by $\vec{P} \propto \vec{e}_{ij} \times (\vec{S}_i \times \vec{S}_j)$. Generally, the direction of $\vec{e}_{ij}$ is along the spin chain and $(\vec{S}_i \times \vec{S}_j)$ is the spin rotation axis. In a proper screw magnet, due to $\vec{e}_{ij} \parallel (\vec{S}_i \times \vec{S}_j)$, the value of *P* is zero. However, for Ba$_9$Fe$_3$Se$_{15}$, the spin chain is slightly tilted by 0.11 ° from the spin rotation axis due to the monoclinic structure with $\beta$=90.11 °[22]. Therefore, according to the prediction by the spin-current model, non-zero *P* can be generated in Ba$_9$Fe$_3$Se$_{15}$, and the direction of *P* should be along the -*b* axis, as can be seen in Fig. 6.

Our first-principles calculations confirm the existence of polarization driven by magnetism. As shown in Fig. 2(b), the magnetic structure is incommensurate and the *c*-axis magnetic lattice parameter is about 143.6 Å, corresponding to about 8 unit cells. It is not possible to carry out the calculation in a complete magnetic lattice, because of the large atomic numbers. As a compromise, to simplify the calculation, we just separate the magnetic lattice into 8 unit cells and only deal with the first three unit cells separately. The periodic boundary condition with one unit cell as a period was adopted. Although it would change the adjacent spin angles when crossing the boundary, it still can help to qualitatively understand the origin of polarization. Our first-principles calculations show that the spin induced polarization of the three different unit cells are 14.79 μC-m$^{-2}$, 13.49 μC-m$^{-2}$ and 13.69 μC-m$^{-2}$, respectively, which are of magnitudes consistent with the experimental results. In addition, when the magnetic moment of each Fe atom undergoes mirror operation, the DM interaction induced polarization would be reversed. Therefore, our calculation results prove that the spin induced polarization is an intrinsic phenomenon of Ba$_9$Fe$_3$Se$_{15}$. To further study the mechanism of spin induced polarization, we calculated the polarizations when varying the $\beta$ value in the monoclinic structure and the results are shown in Fig. S4. It is found that the polarization increases with $\beta - \frac{\pi}{2}$, which agrees qualitatively with the prediction by the spin-current model.

Generally, for the spin induced ferroelectricity, the polarization coincides with the LRSO. However, in Ba$_9$Fe$_3$Se$_{15}$, the polarization temperature is slightly higher than $T_N$. This phenomena should arise from the reduction of dimensions. Ba$_9$Fe$_3$Se$_{15}$ is featured with well-separated quasi 1D spin chains. Above $T_N$ abundant SRSOs associated with the intrachain coupling have been developed within the decoupled spin chains. During the poling process the helicity $(\vec{S}_i \times \vec{S}_j)$ of the helical spin orders is fixed, i.e., the system is in the polarized state. When it is warmed across $T_N$, the 3D LRSO is destroyed, and the spin chains are decoupled. However, the spin orders with moderate length still exist within the decoupled chains and the helicity of the spin order has not been changed yet, which helps to keep the polarized state. Therefore, the polarization can survive above $T_N$ in the quasi 1D helical magnet of Ba$_9$Fe$_3$Se$_{15}$. It should be noted that below $T_N$ the polarized domains are 3D and permeate the whole sample, while above $T_N$ they are confined within the well-separated spin chains.

For the same reason of strong 1D spin chain characteristics, Ba$_9$Fe$_3$Se$_{15}$ exhibits different dielectric behaviors from other spin-induced ferroelectrics in 2D/3D systems. Usually, the ferroelectric polarization generated by LRSO causes an abrupt increase in the dielectric permittivity $\varepsilon_r$, demonstrating a sharp peak and frequency independent behavior. In contrast, in $Ba_9Fe_3Se_{15}$, the frequency dependent dielectric permittivity $\varepsilon_r$ and dielectric loss, *i.e.*, dielectric relaxation behavior was observed. The relaxation can be understood from the perspective of SRSOs which can generate dipole entities. Far above $T_N$, the intrachain spin correlations have already been pre-developed to form SRSOs, which has been suggested by the specific heat measurements. The SRSOs have two different helicities, *i.e.*, the right- and left-handed rotations, thus the generated polarized entities have two equivalent orientations. These polarized entities within the decoupled chains can be oriented by applying AC electric field and contribute to the measured $\varepsilon_r$. When the temperature is cooled down towards $T_N$, the SRSOs gradually grow due to the development of intrachain spin correlations and its dynamic is slowed down, which leads the decrease of $\varepsilon_r$. When the temperature is reduced across $T_N$, the long-range ferroelectric order forms and $\varepsilon_r$ decreases quickly. Thus, for $Ba_9Fe_3Se_{15}$, we cannot observe a sharp peak of $\varepsilon_r$ corresponding to the LRSO transition which usually is seen in 2D/3D spiral magnets. In Fig. 5(a,b), the decrease of $\varepsilon_r$ value and the shift of tan$\delta$ to high temperature with increasing magnetic field possibly imply that magnetic field should enhance the intrachain spin correlations and thus increase the SRSO length. Since the SRSO size and its dynamics are temperature dependent, the response to the external AC electric field should depend on frequency. This situation is analogous to what happens in compositionally disordered $ABO_3$ perovskites, where the random disorders can lead to the formation of dipolar defects[30-37]. If the dipolar defects are in the highly polarizable dielectric host lattice, they can be correlated to form polarized nano-domains and lead to dielectric relaxation under AC electric field. In addition, the domain size is determined by the temperature dependent correlation length of the host lattice and the domains grow with temperature decreasing. If the concentration of the nanodomains is sufficiently high, the polarized domains will permeate the whole lattice and form a long-range ordered ferroelectric state. Here, in $Ba_9Fe_3Se_{15}$, the polarized nanodomains are generated by the SRSOs and the long-range ordered ferroelectric state is realized via the formation of LRSO.

## Conclusions

In summary, the spin induced ferroelectricity was observed in tilted screw magnetic $Ba_9Fe_3Se_{15}$ which is featured with a strong 1D spin-chain structure. Far above $T_N$, SRSOs generate the dipolar entities and respond to the external AC electric field. This causes a frequency dependent dielectric permittivity, $\varepsilon_r$ and dielectric loss, tan$\delta$ as well as magnetoelectric coupling behavior. When decreasing the temperature across $T_N$, the dipolar entities grow and their dynamic responses slow down gradually and permeate the whole lattice to form three-dimensional ferroelectricity at $T_N$. These results support that the novel polarization dynamic in $Ba_9Fe_3Se_{15}$ is due to its quasi-1D character.

## Experiments

$Ba_9Fe_3Se_{15}$ was synthesized by solid state reaction at high pressure and high temperature according to a previous report[22]. Small rod-shaped single crystals (1 mm in length) can be extracted from the polycrystalline sample. The DC magnetic susceptibility was measured with a superconducting quantum interference device (SQUID). A physical property measuring system (PPMS) was used for thermodynamic property measurements, for performing AC susceptibility measurements that were carried out with zero DC background field and an amplitude of 8 Oe.

A sample of 3 mm in diameter and 0.3 mm in thickness was used for the measurements of dielectric permittivity and pyroelectric current $I_p$. The dielectric permittivity was measured using an Agilent-4980A LCR meter attached to the PPMS. The $I_p$ was measured using an electrometer (Keithley 6517B) connected with the PPMS. Before $I_p$ was measured, the sample was electrically poled by applying an electric field $E$ of ±6.6 kV at 20 K. After it was cooled down to 2 K, the polarization $E$ was removed. The sample was short-circuited for half an hour to exclude the contributions other than polarized charges, and then the pyroelectric current $I_p$ was collected when the temperature rose with a ramping rate of 2 K. The electric polarization $P$ was obtained by integrating $I_p$ as a function of time.

The $Ba_9Fe_3Se_{15}$ sample was measured by NPD with neutron wavelength ($\lambda$) of 2.3579 Å on HB-3 Triple-axis spectrometer at the High-Flux Isotope Reactor at the Oak Ridge National Laboratory. The neutron-diffraction data were refined using the FULLPROF software package including the magnetic symmetry analysis with BasiReps.

The first-principles calculations of $Ba_9Fe_3Se_{15}$ have been carried out by using the Vienna Ab initio Simulation Package (VASP) with a plane wave basis set and the projector-augmented wave (PAW) method[38, 39]. The generalized gradient approximation with the Perdew-Burke-Ernzerhof parameterization (GGA-PBE) was used as the exchange-correlation functional [40]. The cutoff energy for plane-wave basis was set to 500 eV and 5×8×5 Monkhorst-Pack k-points was adopted for Brillouin zone sampling. Noncollinear magnetism and spin-orbit coupling were included in all calculations. The Hubbard U = 2.0 eV was taken into account to describe the on-site Coulomb repulsion between Fe-3d electrons. All the structures were optimized until the Hellmann−Feynman forces were below 1 meV/Å, and the convergence threshold of electronic energy was $10^{-6}$ eV. In addition, the spontaneous ferroelectric polarization (Ps) was determined by the Berry phase method[41, 42].

**Acknowledgments**

We greatly appreciate the support of the National Key R&D Program of China, the Natural Science Foundation of China and the Chinese Academy of Sciences (Grant No. 2018YFA0305700, 11974410, 2018YFE0103200, 11921004, 11934017, 11904392, 11534016, 2020YFA0711502, 2017YFA0302900 and XDB33000000). The neutron experiment at the Materials and Life Science Experimental Facility of the J-PARC was performed under a user program (Proposal No. 2019S06). C.-J. K. was supported by the National Research Foundation of Korea (NRF) under Grant No. NRF-2022R1C1C1008200.

**Author contributions**

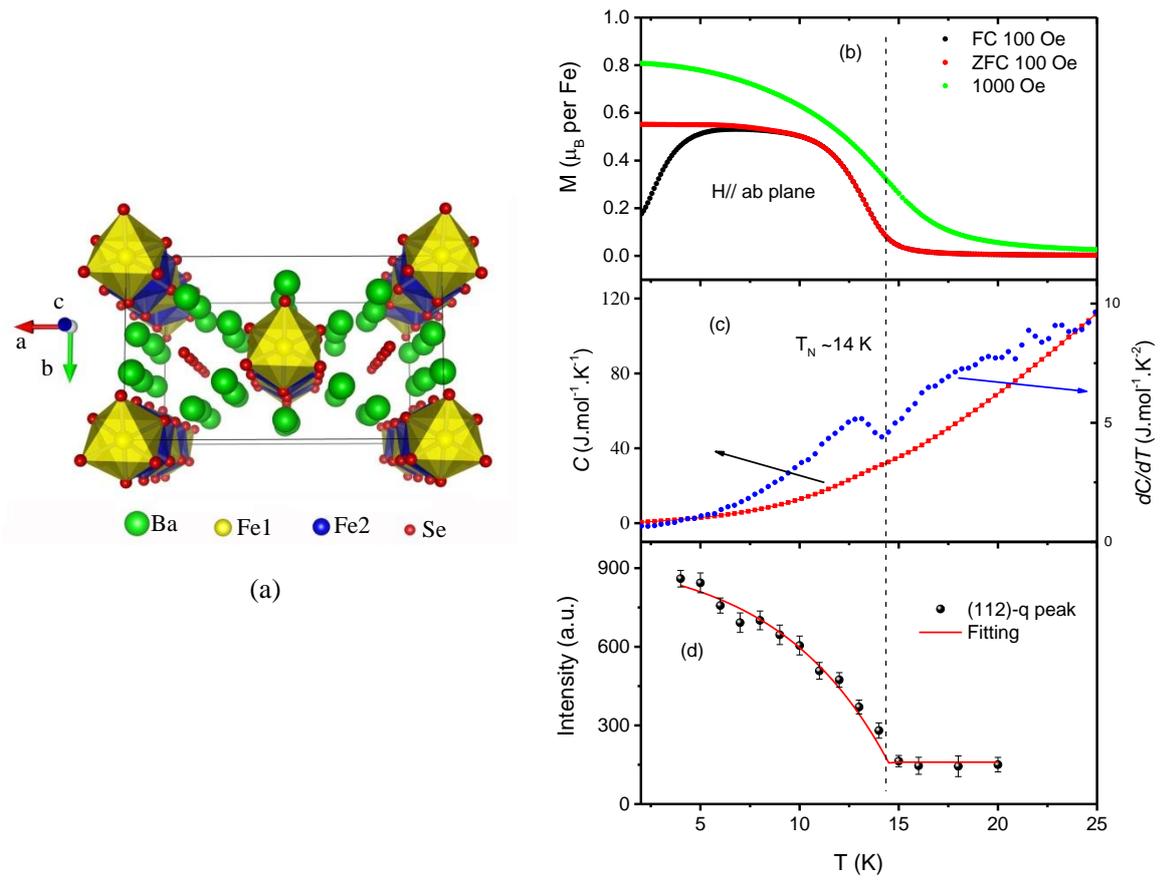

Fig. 1 (a) The sketch of crystal structure of $Ba_9Fe_3Se_{15}$. (b) The temperature dependence of DC magnetization for single crystalline $Ba_9Fe_3Se_{15}$ sample under different magnetic fields with $H//ab$-plane. (c) The specific heat at 0 T and the temperature derivative of specific heat. (d) The temperature dependence of integrated intensity for the magnetic Bragg peak (1 1 2)-q.

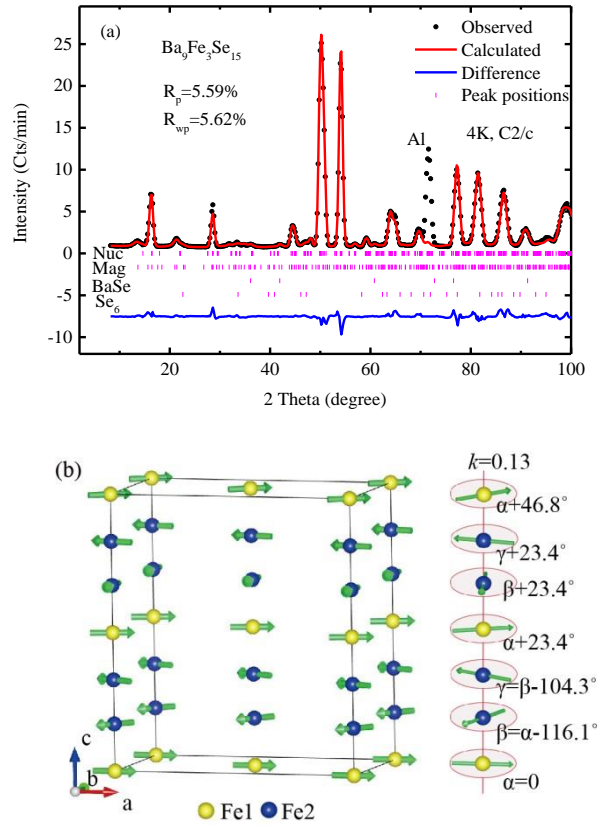

Fig. 2 (a) The Rietveld-fit of the magnetic intensities measured by powder neutron diffraction by the HB-3 Spectrometer ($\lambda = 2.3579$ Å) at 4 K for $Ba_9Fe_3Se_{15}$ sample. (b) The sketch of the magnetic structure, showing a spin spiral configuration.

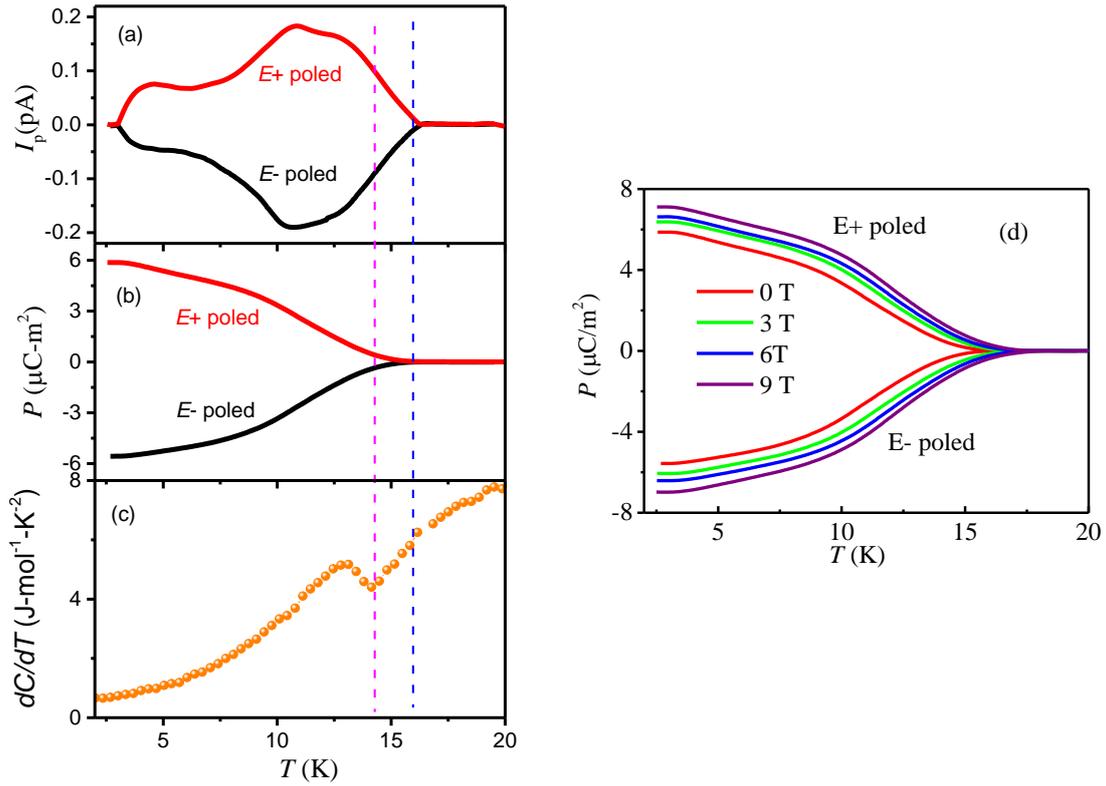

Fig. 3. (a) The temperature dependence of pyroelectric current $I_p$ poling from 20 K to 2 K, and (b) the electric polarization $P$ obtained via the integration of $I_p$ for $Ba_9Fe_3Se_{15}$. (c) The temperature derivative of specific heat is re-plotted. The violet dashed line represents the long-range spin ordering transition temperature while the blue one for the temperature where the non-zero polarization occurs. (d) The electric polarization under different magnetic fields.

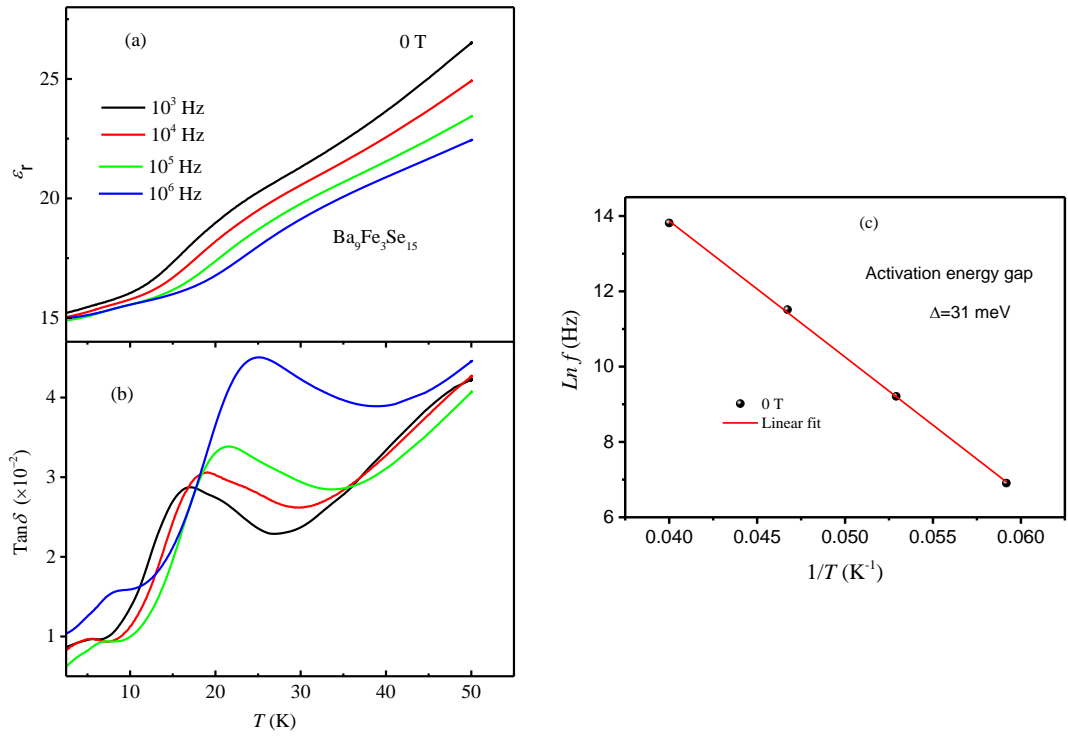

Fig. 4. (a) The temperature dependence of dielectric permittivity $\varepsilon_r$, and (b) dielectric loss tan$\delta$ under different frequencies and zero magnetic field for $Ba_9Fe_3Se_{15}$. (c) The relationship between the frequency and the inverse temperature of the dielectric loss peak. The red line is the linear fit.

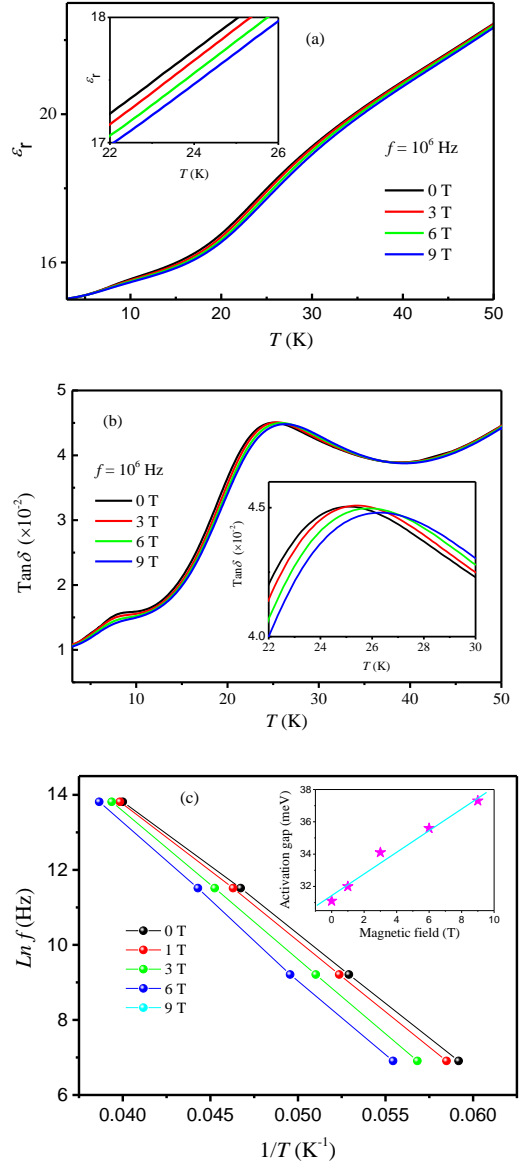

Fig. 5. (a) The temperature dependence of dielectric permittivity $\varepsilon_r$, and (b) dielectric loss tan$\delta$ under different magnetic field and fixed $f = 10^6$ Hz for $Ba_9Fe_3Se_{15}$. The insets of (a) and (b) show their enlarged view, respectively. (c) The relationship between $Ln(f)$ and the peak temperature of tan$\delta$ for different magnetic field. The inset is the activation energy gap versus magnetic field.

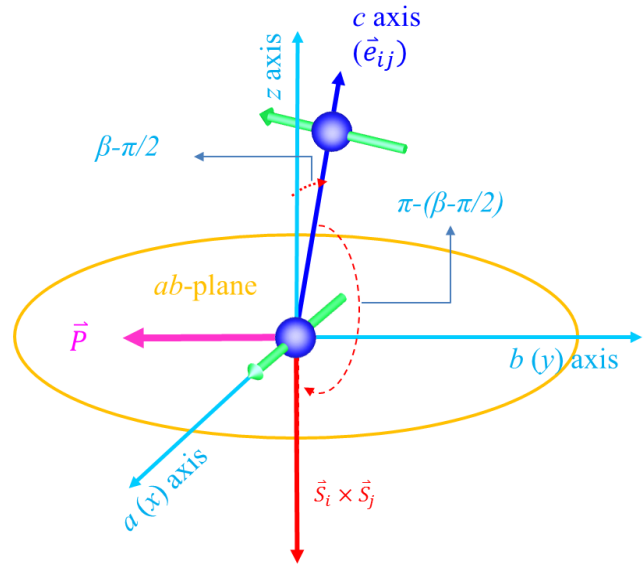

Fig. 6 The sketch of the spin induced polarization within the spin current model for Ba$_9$Fe$_3$Se$_{15}$ with tilted helical spin arrangement. Due to the monoclinic structure the spin rotation axis $(\vec{S}_i \times \vec{S}_j)$ and the $c$ axis (spin chain direction) form the angle of π-(β-π/2). The generated $P$ is along the b axis.

# Supplementary materials

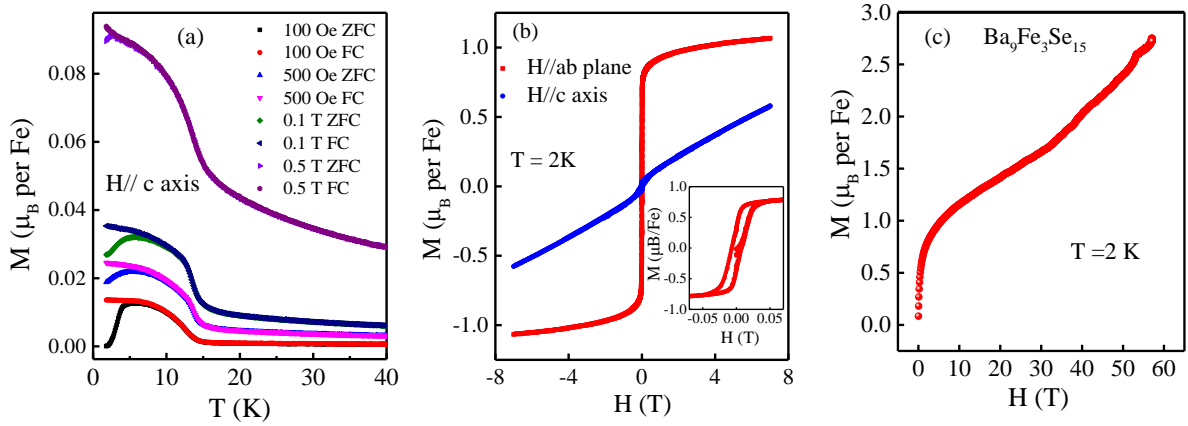

Fig. S1 (a) The magnetization measured under different fields with the applying magnetic field H//c axis. (b) The magnetic hysteresis loops measured at 2 K with *H//ab*-plane and *H//c*-axis, respectively. The inset is the enlarged view to clearly show the coercive force. (c) The magnetization versus magnetic field for polycrystalline $Ba_9Fe_3Se_{15}$ sample measured at 2 K within 60 T.

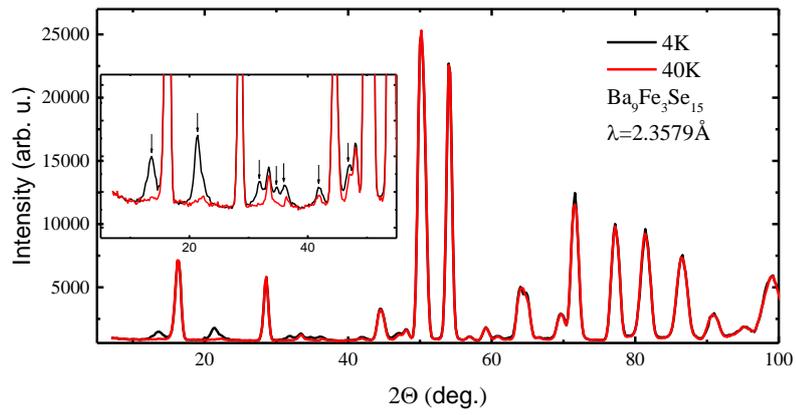

Fig. S2 The neutron diffraction data collected at 4 K and 40 K for $Ba_9Fe_3Se_{15}$. The inset is the enlarged view to show the new magnetic peak collected at 4 K marked by the arrows.

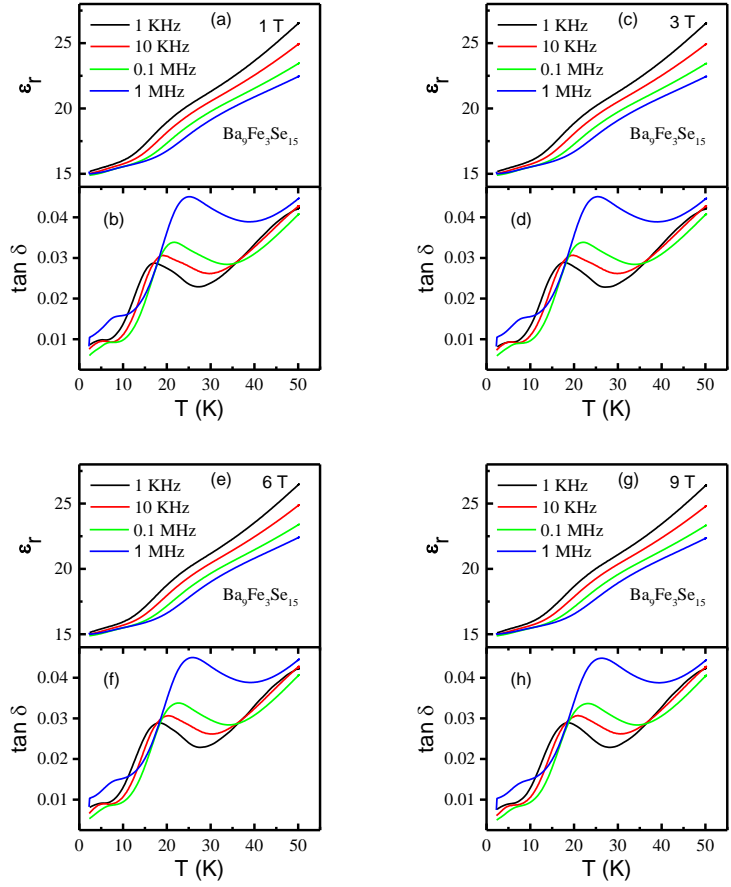

Fig. 3S(a-h) The temperature dependence of dielectric permittivity $\varepsilon_r$ as well as dielectric loss tan$\delta$ measured under different frequencies and under magnetic field of 1 T, 3 T, 6 T and 9 T, respectively.

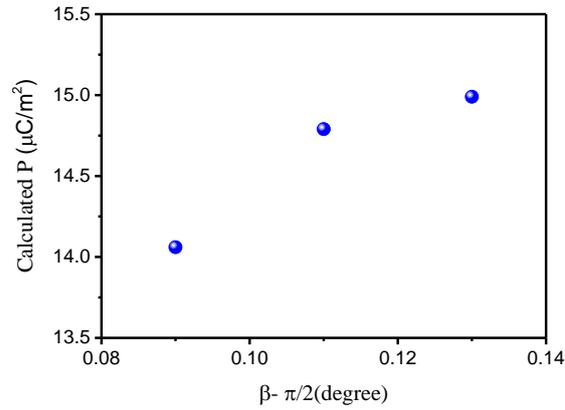

Fig. S4. The calculated polarization dependent on angle $\theta = \pi - (\beta - \frac{\pi}{2})$ for $Ba_9Fe_3Se_{15}$.